%% file: 00paper.tex
\documentclass[sigconf]{acmart}

\usepackage{booktabs}
\usepackage{algorithm}
\usepackage{algpseudocode}
\usepackage{bbold}
\usepackage{enumitem}
\usepackage{subfigure}
\usepackage{colortbl}

\usepackage{setspace}
\usepackage{soul}

\usepackage{graphicx}
\newcommand{\todo}[1]{\textcolor{red}{#1}}

\newtheoremstyle{mydef}
{2ex}
{2ex}
{\itshape}
{}
{\scshape}
{: }
{0.5em}
{}
\theoremstyle{mydef}

\newcommand{\myparagraph}[1]{\paragraph*{\hspace*{-\parindent}\normalsize{\bf{#1}}}}

\begin{document}

\copyrightyear{2020}
\acmYear{2020}
\setcopyright{acmcopyright}\acmConference[SIGIR '20]{Proceedings of the 43rd International ACM SIGIR Conference on Research and Development in Information Retrieval}{July 25--30, 2020}{Virtual Event, China}
\acmBooktitle{Proceedings of the 43rd International ACM SIGIR Conference on Research and Development in Information Retrieval (SIGIR '20), July 25--30, 2020, Virtual Event, China}
\acmPrice{15.00}
\acmDOI{10.1145/3397271.3401205}
\acmISBN{978-1-4503-8016-4/20/07}



\title{Summarizing and Exploring Tabular Data in  Conversational Search}
\author{Shuo Zhang}
\authornote{The first two authors contributed equally.}
\affiliation{%
  \institution{Bloomberg}
  \city{London}
  \country{United Kingdom}
}
\email{szhang611@bloomberg.net}
\author{Zhuyun Dai$^*$}
\affiliation{%
  \institution{Carnegie Mellon University}
  \city{Pittsburgh}
  \country{USA}
}
\email{zhuyund@cs.cmu.edu}
\author{Krisztian Balog}
\affiliation{%
  \institution{University of Stavanger}
  \city{Stavanger}
  \country{Norway}
}
\email{krisztian.balog@uis.no}
\author{Jamie Callan}
\affiliation{%
  \institution{Carnegie Mellon University}
  \city{Pittsburgh}
  \country{USA}
}
\email{callan@cs.cmu.edu}

\renewcommand{\shortauthors}{Zhang et al.}

\begin{abstract}
Tabular data provide answers to a significant portion of search queries. However, reciting an entire result table is impractical in conversational search systems.
We propose to generate natural language summaries as answers to describe the complex information contained in a table.
Through crowdsourcing experiments, we build a new conversation-oriented, open-domain table summarization dataset. It includes annotated table summaries, which not only answer questions but also help people explore other information in the table. 
We utilize this dataset to develop automatic table summarization systems as SOTA  baselines. 
Based on the experimental results, we identify challenges and point out future research directions that this resource will support.

\end{abstract}

%


\maketitle

\input{00paper-01}

\input{00paper-02}

\input{00paper-03}

\input{00paper-04}

\myparagraph{Acknowledgments}

This work was partially supported by the National Science Foundation (NSF) grant IIS-1815528. 

\bibliographystyle{ACM-Reference-Format}
\bibliography{00paper}

\end{document}

%% file: 00paper-01.tex
\vspace*{-0.5\baselineskip}
\section{Introduction}

Many search queries are seeking a set of items and their attributes, e.g., ``highest-grossing movies'' or ``best places to travel in Christmas.''
In a typical information retrieval system, such search results can be presented as a list or table.
In browser-based search, approximately 10\% of QA queries are answered by tabular data~\citep{Zhang:2020:WTE}.
While structured data is available in large quantities, presenting search results in a tabular format is challenging in conversational search systems (e.g., Siri, Alexa, and Google Assistant).
First, tables are rich in structure while low in natural language content.  Reading out a table line by line will lead to a poor user experience.  Second, tables can be quite large. They cannot be presented effectively in chat or voice-only user interfaces.
%

\begin{figure}[t]
   \centering
   \includegraphics[width=0.4\textwidth]{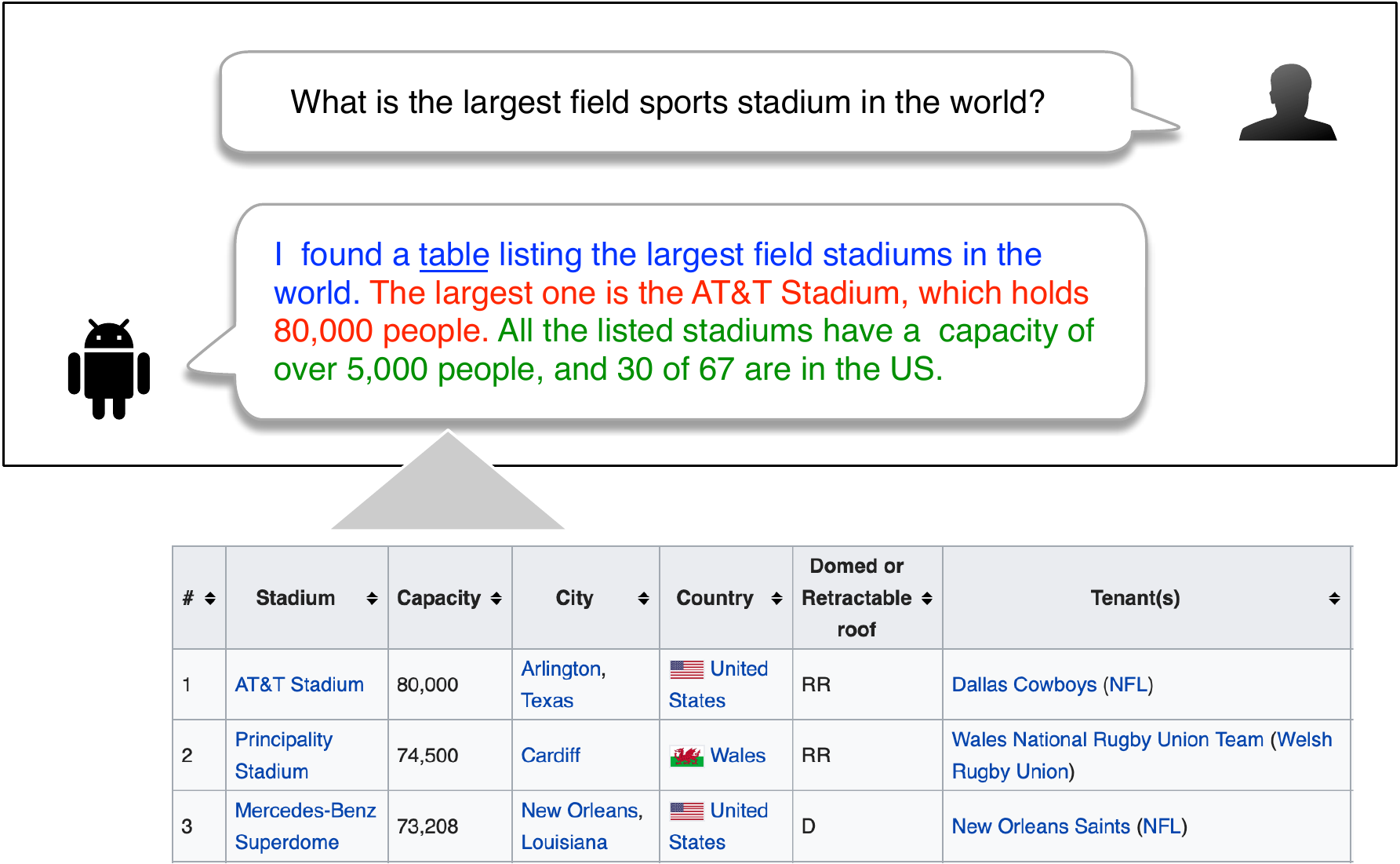} 
   \caption{Table summarization in conversational search. The summary of the result table includes a leading sentence describing the table (red), answer to the question (blue), and a sentence helping to explore the table further (green).} 
   \vspace*{-0.5\baselineskip}
\label{fig:task}
    \vspace{-1\baselineskip}
\end{figure}

In this work, we aim to bridge this gap by facilitating natural language summarization of tables in conversational search.
Specifically, we wish to leverage the interactive nature of multi-turn conversational systems. Instead of telling users everything at once, we seek to present a concise \emph{summary} as well as give users information that helps to further \emph{explore} the table's content.  That is, we want the system to \emph{drive the conversation}---provide clues that aid users in what they could ask next.
We illustrate the idea in Fig~\ref{fig:task}.

There exist structure-based as well as content-based methods for table summarization.
The former can rely on a single predefined schema~\citep{Jain:2018:AMH}, automatically generated attribute value taxonomies~\citep{Ienco:2013:KFT}, or patterns succinctly summarizing the tables~\citep{Chen:2013:TSF}. 
The latter can utilize value lattices~\citep{Selcuk:2009:ASS} or table rows~\citep{Ming:2000:TSA} by presenting partial content. 
However, these methods assume a decomposable table structure and are often limited to predefined schemes, and are not conditioned on the conversation context that triggered the table. 

It is an open research question what types of summaries can help people to explore the table in conversations. 
 We propose a table summary with the following traits. 
 First, the summary should contain the \emph{answer to the user's question} asked in the last conversation turn.  
For example, ``the world's largest field stadium is the AT\&T stadium, and it can hold 80,000 people.''
Second, the summary should let the user know what is inside the table, so that they can further explore its contents.  Thus, the summary may provide an \emph{overview of the table}, e.g., ``the table lists the largest field sports stadiums in the world,'' or \emph{additional information} from the table, e.g, ``30 of 67 are from the US.'' In this way, the user can keep exploring the table by asking questions like ``What is the second largest one?''   

As a first step towards the development of automatic approaches, we create a test collection for this task.  
Specifically, we use the above traits to define crowdsourcing tasks, and collect manually-written table summaries along with corresponding relevance assessments.  We also compares existing state-of-the-art natural language generation models on our dataset to gain further insights.

In summary, the main contributions of this work are:
(1) introduction of the task of table summarization in conversational search;
(2) creation of a test collection using crowdsourcing, which is made publicly available;\footnote{https://github.com/iai-group/sigir2020-tablesum}
(3) investigation of how general-purpose abstractive summarization methods perform as baselines;
(4) analysis of the baseline results and identification of future directions.

%% file: 00paper-02.tex
\vspace*{-0.5\baselineskip}
\section{Creating a Test Collection} 

Our objective is to create a test collection to study table summerization in a conversational setting.
Specifically, given a user query $q$ in a conversation and a result table $T$, we aim to create a summary $S$ of this table that can help answer the query.
We describe the sampling of queries and tables in Sect.~\ref{sec:sub:tbl} and detail the crowdsourcing experiments we carried out to collect summaries and annotations in Sects.~\ref{sec:sub:cs} and~\ref{sec:sub:ra}, respectively.


\vspace*{-0.5\baselineskip}
\subsection{Queries and Tables}
\label{sec:sub:tbl}

Our test collection comprises of 200 tables, each with a corresponding question as the conversation context. That is, $\langle q, T\rangle$ pairs constitute the input to summary generation. 

The tables are randomly selected from the WikiTableQuestions dataset~\cite{pasupat2015compositional}, which has formerly been used for semantic parsing on tables. 
This dataset contains 2108 Wikipedia tables on a large variety of topics, and over 22k questions for querying these tables.  While this dataset can be reused for our task, it does not provide table summaries.
We require each selected table to have at least six rows and four columns, because tables smaller than this may be displayed in their entirety, without needing summarization.
We manually identify the type of each table. There are 64 Sport, 33 Place, 27 Music, 16 Film, 12 Culture, 11 Traffic, 8 Product, and 30 Other tables (including Politics, TV series, Award, and Company). 

The questions are either sampled from the WikiTableQuestions dataset (45 questions), or written by two of the authors (155 questions). The former are mostly fact-checking questions, e.g., ``How many points did Toronto have more than Montreal in their first game?'' 
The latter contains both
fact-checking questions and more open-ended ones, e.g., ``tell me about the car models made by Ford.''


\vspace*{-0.5\baselineskip}
\subsection{Collecting Candidate Summaries}
\label{sec:sub:cs}

We aim to collect summaries that are suitable in a conversational setting.  To achieve this goal, we designed a crowd-sourcing task that mimics a real conversation with a friend.  The instructions were given such that the desired traits for table summaries are emphasized: \emph{relevant to the question} and provides \emph{clues for exploration}.

Specifically, the instructions given for the crowdsourcing task are as follows:
``Image talking to a friend on the phone. Your friend asks you a question, and you find the following table on the Web.  Remember that your friend cannot see the table.  Your goal is to let your friend to capture essential information in the table related to the question.  Your summary should be short but comprehensive.  Try to describe several rows or columns that you find interesting.''

\begin{figure}[t]
   \centering
   \vspace*{-0.75\baselineskip}
   \includegraphics[width=0.5\textwidth]{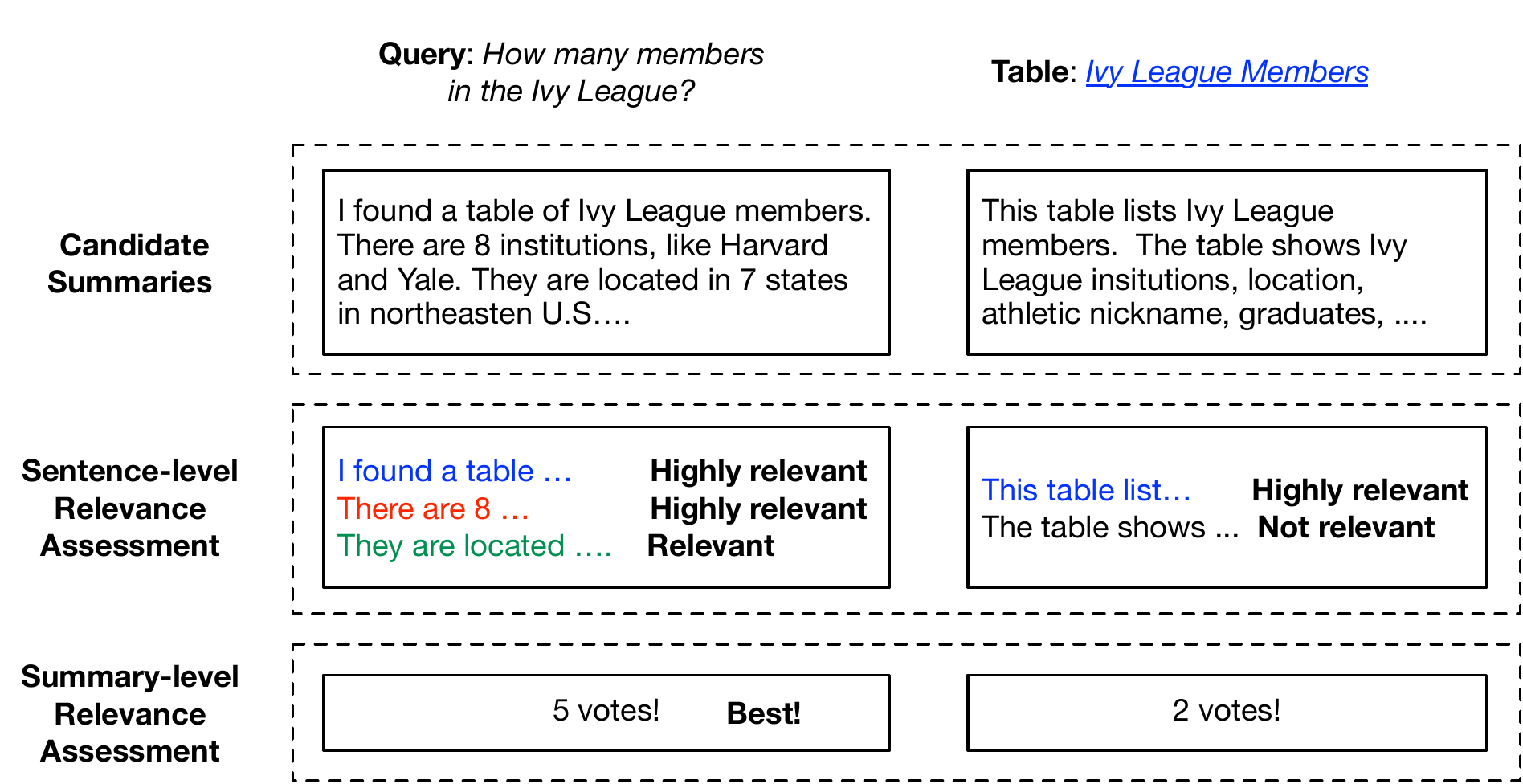} 
   \caption{Illustration of our crowdsourcing workflow.}
\vspace*{-1\baselineskip}
\label{fig:cw_task}
\end{figure}

As shown in Fig~\ref{fig:cw_task}, we show the question, the title of the table, and its Wikipedia page link to the crowd worker. The worker needs to click on the link, read the table, and write the summary in a text input box that has a character counter. We limit the summary to be around in 30-50 words (150-250 characters).\footnote{This is based on that (i) the short text message (SMS) limit is 160 chars, and we want the summary to fit approximately within that limit, and (ii) in spoken English, people speak 125 words per minute, and we want the summary to be 15-30 seconds long.} 
For any $\langle q, T\rangle$ pair, we invited five workers and suggested them spend at least 5 minutes in writing the summary. The average time per assignment reported from Amazon Mechanical Turk was 19 minutes 36 seconds.
In the end, we collected $200 \times 5 = 1000$ candidate summaries.

\vspace*{-0.5\baselineskip}
\subsection{Quality Assessment}
\label{sec:sub:ra}

Given the set $S$ of five summaries collected in Sect.~\ref{sec:sub:cs}, we aim to find the best one by labeling them with relevance through a second set of crowdsourcing experiments.
Specifically, ground truth relevance labels are obtained for each sentence (Sect.~\ref{sec:sub:sub:sl}) as well as for the entire summary (Sect.~\ref{sec:sub:sub:suml}). 

\subsubsection{Sentence Level}
\label{sec:sub:sub:sl}
We collect sentence-level relevance assessments by employing three judges (see the middle block in Fig.~\ref{fig:cw_task} for the illustration of this step). 
Each summary was split into sentences. 
In total there are 3459 sentences, i.e., 3.5 sentences per summary on average.
We show each sentence to the judges, with its previous and next sentences provided as context. 
Sentences were judged on a three-point scale: highly relevant, relevant, and non-relevant.
The annotators were situated in a scenario where they need to write a short summary.
This summary should help their friend to capture the information contained in this table easily, and should provide a fact as an answer to the question. 
Specifically, they were given the following guidelines: (i)
a sentence is \textbf{highly relevant} if it tells the table topic or 
provides a fact regarding the friend's question;
(ii) a sentence is \textbf{relevant} if 
it is not directly relevant to the friend's question, but helps to figure out what the table contains for those who can not see it;
and (iii) a sentence is \textbf{not relevant} if it is not about the table or is unrelated to the question.

\subsubsection{Summary Level}
\label{sec:sub:sub:suml}

Summary-level assessments were collected by showing the judges all candidate summaries for a table, and asking them to select the \emph{best} one (see the bottom block in Fig.~\ref{fig:cw_task} for the illustration of this step).
We define quality guidelines based on the following three aspects: (i) language quality: the summary should be concise and easy to read; (ii) relevance: the summary should provide a fact regarding the friend's question; and (iii) able to drive conversations: the summary should be able to attract the reader's interests, and intrigue the reader to ask follow-up questions. For example, the summary may provide an overview of the table, or show some highlights from the table that may not directly relate to the question.

For each table, the five collected summaries (cf. Sect.~\ref{sec:sub:cs}), together with the question, were shown to 7 judges.\footnote{The number of judges were based on our budget.} The summaries were organized in a list; the order of summaries was randomly shuffled each time to eliminate potential rank bias. Neither the table itself nor the Wikipedia URL were given to the judge, to mimic the real use case where the user only sees/hears the summary but cannot see the table. 
The judges were asked to pick the highest quality summary from the five based on the above guidelines. The one with the most votes becomes the ground truth summary for the table.

Among the 200 top-voted summaries, the numbers of summaries that obtain 5, 4, 3, and 2 votes are 5, 34, 115, and 46, respectively. In other words, for 154/200 (77\%) of the tables, at least 3 judges agreed on which summary was the best.
By analyzing the top-voted summaries, we found that the judges tend to prefer summaries that are longer and use a diverse vocabulary (unique words). 
On average, the top-voted summaries have $53.1$ words and $41.2$ unique words; for comparison, an average summary has 46.2 words and 36.9 unique words, while the least-voted summary has 40.8 words and 33.3 unique words. Additionally, we found that the top-voted summary uses more numbers. This, to a certain extent, shows that these summaries contain more factual information. These findings are aligned with our guidelines, which require the summary to provide comprehensive information about the table to intrigue follow-up dialogue. 
%



%% file: 00paper-03.tex
\vspace*{-0.25\baselineskip}
\section{Experiments} 
\label{sec:ats}

In this section, we compare the performance
of several state-of-the-art natural language generation models~\citep{Gu:2016:ICM,Vaswani:2017:AAY,raffel2019t5} on our dataset. Our aim is to understand the extent to which table summaries can be generated automatically using current state-of-the-art approaches, and to gain insights into what the challenges of this task are.

%



%
%
\vspace*{-0.5\baselineskip}
\subsection{Methods for Comparison}
\label{sec:ats:meth}




%

This section first introduces how we represent tables for the neural language generation models. Next, it introduces three stage-of-the-art models for comparison.

\textbf{Representating Tables as Text Sequence}. Most of the current state-of-the-art natural language generation models  expect the input to be a 1-dimensional text sequence.  Inspired by ~\citet{hancock2019tabletitle}, we flatten tables into a sequence of words, and use a ``key:value'' format to preserve the structure. 
Specifically, we take one row at a time, and pair the cells with the corresponding column header. 
For example, the cell value ``164,119'' in the column ``Sales'' generates ``Sales:164,119.''  Next, we concatenate the header-cell pairs in the same row using commas, and  concatenate the rows  to represent the entire table. Row indexes are listed at the beginning of the corresponding rows.
Finally, we add the page title and table caption to provide context. We also add the total number of rows to the flattened table, because many manual summaries start with ``there are in total X rows.''
The page title, table caption, and the total number of rows are also represented in ``key: value'' format, where the key is ``PageTitle,'' ``Caption,'' or ``TotalRows,'' and the value is the corresponding text. During training, the models are expected to understand the meanings of these keys and learn to use their values to generate summaries.

We consider three  neural natural language generation models to generate summaries based on the flattened tables. We take these methods as state-of-the-arts baselines that others can reproduce and use to test their own solutions.

\textbf{CopyNet}~\citep{Gu:2016:ICM} is an LSTM-based encoder-decoder model that incorporates the copying mechanism. 
During training, the model encodes the input table using a layer of bidirectional LSTM and tries to decode it into the human-written summary. 
The copying mechanism can choose sub-sequences in the input and put them at proper places in the output. 
During testing, the model only sees the input table and automatically generates a text summary. 

\textbf{GPT-2}~\cite{radford2019language}  is a large language generation model pre-trained over 40GB of text data crawled from the Web.  
Unlike CopyNet, which must be learned from scratch, GPT-2 already learns general language patterns and needs less task-specific training data. 
It is a good fit for our task as we only collected summaries for 200 tables. 
GPT-2's model consists of the decoder part of the Transformer~\citep{Vaswani:2017:AAY}, and was trained with a causal language modeling objective. 
Since there is no encoder in GPT-2, the model does not have a clear separation between the input table and the target summary. 
Therefore, we concatenate the table and the summary, separated by a special ``\#summary\#'' token. We train GPT-2 on this text input to predict every token in the sequence conditioned on its previous tokens. 
For testing, we feed GPT-2 with the original table + ``\#summary'' and let GPT-2 predict the following tokens.

\textbf{Text-to-Text Transfer Transformer (T5) }~\cite{raffel2019t5} is another large, pre-trained language generation model. 
T5 employs both the encoder and the decoder part of the Transformer, so that it can be trained and tested with standard (table, summarization) pairs as used for CopyNet. Compared to GPT-2, T5 is pre-trained on an even larger corpus of 645GB text data using a cleaned subset of Common Crawl. T5 achieved state-of-the-art results on several summarization benchmarks~\cite{raffel2019t5}. 


\textbf{Implementation Details}. We used the AllenNLP~\cite{Gardner2017AllenNLP} implementation of CopyNet, with 300-dimensional GloVe embeddings, one-layer bidirectional LSTM for encoding, and one-layer LSTM for decoding. 
GPT-2 used the Huggingface~\cite{Wolf2019HuggingFacesTS} implementation. 
The small version of GPT-2 was used due to GPU memory limitation.\footnote{GPU type is Nvidia GeForce RTX 2080 Ti with 11GB memory} Training follows the hyper-parameters recommended by~\citet{Wolf2019HuggingFacesTS}. 
T5 used the implementations provided by the authors.\footnote{https://github.com/google-research/text-to-text-transfer-transformer} It is trained on Google TPU using the default hyperparameters. For all models, we only used the first 256 tokens from the table.

The training process was through 5-fold cross-validation. Each fold used $160$ tables for training and $40$ tables for testing. To test the model, we compare the automatically generated summary with the best summary that received the most votes from the judges.  We use ROUGE and BLEU as our evaluation metrics. 
During training, we used any summary that had non-zero votes. 
For each table, $3.6$ out of the $5$ candidate summaries get at least one vote.
This brings more training data than just using the top-voted summary, and was shown to lead to slightly better experiential performance.


\vspace*{-0.5\baselineskip}
\subsection{Experimental Results and Discussion}
\label{sec:ats:er}

\begin{table}[t]
  \centering
  \caption{Evaluation results of three automatic summarization methods. We compare GPT-2 against CopyNet and T5 against GPT-2 for statistical significance testing. $^\ddag$ denotes significance at the 0.005 level.}
  \vspace*{-0.5\baselineskip}
  \begin{tabular}{ l  llll }
    \toprule 
	\textbf{Method} &\textbf{ROUGR-L} & \textbf{ROUGE-1} & \textbf{ROUGE-2}  & \textbf{BLEU} \\
	\midrule
CopyNet & 0.030 & 0.041 & 0.012 & 0.80  \\
GPT-2 & 0.200$^\ddag$ & 0.272$^\ddag$ & 0.073$^\ddag$ & 5.35$^\ddag$ \\
T5 & \textbf{0.276$^\ddag$} & \textbf{0.362$^\ddag$} & \textbf{0.143$^\ddag$} & \textbf{10.43$^\ddag$}  \\

    \bottomrule
  \end{tabular}
  \label{tbl:results1}
  \vspace*{-0.5\baselineskip}
\end{table}
%
%

%
The evaluation results of three models are listed in  Table~\ref{tbl:results1}.
Among the three models, CopyNet achieves the weakest performance on all metrics. CopyNet is not pre-trained and needs much more training examples than our 160 tables.  On the other hand, the two pre-trained models, GPT-2 and T5, can generate reasonable summaries when trained on the limited amount of examples.
Both of them obtain substantial and statistically significant improvements compared to CopyNet.
T5 is the best model for this task.



 Table~\ref{tbl:run-exam} shows a T5-generated summary for the table ``List of number-one albums of 2012 (Finland)''.
It is seen that the machine-generated summary is in fluent natural language. The model is also capable of picking valuable attributes, e.g., the ``Album'' and ``Artists'' columns, while ignoring less valuable attributes, e.g., the ``Reference'' column.
However, we found several mistakes in the machine-generated summaries when manually comparing the human and machine-generated texts. We here conduct an error analysis on all tables and classify the most frequent errors into the following categories:
\begin{itemize}
    \item \emph{Wrong Quantity} (found in 84\% summaries): 
    T5 was prone to making errors in mathematical calculations.  For example, given the table ``Indiana Mr. Basketball Award Winners,''\footnote{\url{https://en.wikipedia.org/wiki/Indiana\_Mr.\_Basketball\#Award\_winners}} the generated summary mentions ``Purdue won 4 consecutive in the 1960s,'' but in fact Purdue only won 3 consecutive.  
    \item \emph{Wrong Reference} (found in 38\% summaries): Some summaries fail to refer to the right attribute. For example, the generated summary mentions
    ``the highest mountain in France is Mont Blanc, at 4,810 meters...'' for table ``List of Alpine peaks by prominence.''\footnote{\url{https://en.wikipedia.org/wiki/List\_of\_Alpine\_peaks\_by\_prominence}} This summary is based on the ``Elevation'' column of the table, but the question asked for ``Prominence.'' 
\end{itemize}

\noindent
In summary, current state-of-the-art text generation models are capable of generating natural language summaries from tables even with a small amount of training examples. However, they do not work well for knowledge reasoning or question answering when complex calculations are involved. 
To address the above problems, a possible research direction for future work is to combine semantic parsing with natural language generation methods. 

%% file: 00paper-04.tex
%
\begin{table}[!t]
  \centering
  \caption{A manually-written summary and a summary generated by T5. Both manage to answer the question correctly and provide extra information contained in the table. 
  }
    \vspace*{-0.5\baselineskip}
\footnotesize
  \begin{tabular}{p{1.5cm}p{6.3cm}}
    \toprule
    Question $q$ & What album had the most sales in 2012 in Finland?\\
    \midrule
    Result table $T$ &  List of number-one albums of 2012 (Finland)\footnote{https://en.wikipedia.org/wiki/List\_of\_number-one\_albums\_of\_2012\_(Finland)}\\
    \midrule
     Manual & I saw a table showing sales figures of around 10 albums of 2012. Of these ``vain elamaa'' by various artists ranked highest regarding sales. It is performed by various artists. The second rank is held by ``koodi'' by robin.
     \\
      T5-Generated & I found a table of the top 10 albums of 2012. The most sold album was ``vain elämä'' by various artists. It sold 164,119 copies. The next highest album was ``koodi'' by robin. \\
    \bottomrule
  \end{tabular}
  \label{tbl:run-exam}
  \vspace*{-0.5\baselineskip}
\end{table}

\section{Conclusion and Future Directions}
\label{sec:cfd}
\vspace*{-0.25\baselineskip}

In this paper, we have introduced the task of table summarization in conversational search and developed a test collection using crowdsourcing. 
The test collection consists of 200 questions, each with a corresponding result table.  Each table comes with five manually created candidate summaries, along with both sentence-level and summary-level quality assessments.
We have employed three neural language generation models as SOTA baselines, performed an experimental comparison of them, and identified two main classes of errors made by these methods.

This paper represents an important first step towards tabular data presentation in conversational search, where the focus was on developing a benchmark test collection.  For baselines, we have focused exclusively on the newest generation of abstractive summarization methods.  
Based on the dataset, it would be interesting to test how more traditional summarization methods~\citep{Jain:2018:AMH,Ienco:2013:KFT,Chen:2013:TSF,Selcuk:2009:ASS} perform on this task.  Additionally, it can be used to design novel approaches that consider the unique characteristics of this task. 


%% file: 00paper.bbl

\begin{thebibliography}{14}


\ifx \showCODEN    \undefined \def \showCODEN     #1{\unskip}     \fi
\ifx \showDOI      \undefined \def \showDOI       #1{#1}\fi
\ifx \showISBNx    \undefined \def \showISBNx     #1{\unskip}     \fi
\ifx \showISBNxiii \undefined \def \showISBNxiii  #1{\unskip}     \fi
\ifx \showISSN     \undefined \def \showISSN      #1{\unskip}     \fi
\ifx \showLCCN     \undefined \def \showLCCN      #1{\unskip}     \fi
\ifx \shownote     \undefined \def \shownote      #1{#1}          \fi
\ifx \showarticletitle \undefined \def \showarticletitle #1{#1}   \fi
\ifx \showURL      \undefined \def \showURL       {\relax}        \fi
\providecommand\bibfield[2]{#2}
\providecommand\bibinfo[2]{#2}
\providecommand\natexlab[1]{#1}
\providecommand\showeprint[2][]{arXiv:#2}

\bibitem[\protect\citeauthoryear{Chen, Pan, Faloutsos, and Papadimitriou}{Chen
  et~al\mbox{.}}{2013}]%
        {Chen:2013:TSF}
\bibfield{author}{\bibinfo{person}{Jieying Chen}, \bibinfo{person}{Jia-Yu Pan},
  \bibinfo{person}{Christos Faloutsos}, {and} \bibinfo{person}{Spiros
  Papadimitriou}.} \bibinfo{year}{2013}\natexlab{}.
\newblock \showarticletitle{TSum: Fast, Principled Table Summarization}. In
  \bibinfo{booktitle}{\emph{Proc. of ADKDD '13}}.
\newblock


\bibitem[\protect\citeauthoryear{Gardner, Grus, Neumann, Tafjord, Dasigi, Liu,
  Peters, Schmitz, and Zettlemoyer}{Gardner et~al\mbox{.}}{2018}]%
        {Gardner2017AllenNLP}
\bibfield{author}{\bibinfo{person}{Matt Gardner}, \bibinfo{person}{Joel Grus},
  \bibinfo{person}{Mark Neumann}, \bibinfo{person}{Oyvind Tafjord},
  \bibinfo{person}{Pradeep Dasigi}, \bibinfo{person}{Nelson~F. Liu},
  \bibinfo{person}{Matthew Peters}, \bibinfo{person}{Michael Schmitz}, {and}
  \bibinfo{person}{Luke Zettlemoyer}.} \bibinfo{year}{2018}\natexlab{}.
\newblock \showarticletitle{{A}llen{NLP}: A Deep Semantic Natural Language
  Processing Platform}. In \bibinfo{booktitle}{\emph{Proc. of Workshop for
  {NLP} Open Source Software ({NLP}-{OSS})}}.
\newblock


\bibitem[\protect\citeauthoryear{Gu, Lu, Li, and Li}{Gu et~al\mbox{.}}{2016}]%
        {Gu:2016:ICM}
\bibfield{author}{\bibinfo{person}{Jiatao Gu}, \bibinfo{person}{Zhengdong Lu},
  \bibinfo{person}{Hang Li}, {and} \bibinfo{person}{Victor~O.K. Li}.}
  \bibinfo{year}{2016}\natexlab{}.
\newblock \showarticletitle{Incorporating Copying Mechanism in
  Sequence-to-Sequence Learning}. In \bibinfo{booktitle}{\emph{Proc. of ACL
  '16}}.
\newblock


\bibitem[\protect\citeauthoryear{Hancock, Lee, and Yu}{Hancock
  et~al\mbox{.}}{2019}]%
        {hancock2019tabletitle}
\bibfield{author}{\bibinfo{person}{Braden Hancock}, \bibinfo{person}{Hongrae
  Lee}, {and} \bibinfo{person}{Cong Yu}.} \bibinfo{year}{2019}\natexlab{}.
\newblock \showarticletitle{Generating Titles for Web Tables}. In
  \bibinfo{booktitle}{\emph{Proc. of WWW '19}}.
\newblock


\bibitem[\protect\citeauthoryear{Ienco, Pitarch, Poncelet, and Teisseire}{Ienco
  et~al\mbox{.}}{2013}]%
        {Ienco:2013:KFT}
\bibfield{author}{\bibinfo{person}{Dino Ienco}, \bibinfo{person}{Yoann
  Pitarch}, \bibinfo{person}{Pascal Poncelet}, {and}
  \bibinfo{person}{Maguelonne Teisseire}.} \bibinfo{year}{2013}\natexlab{}.
\newblock \showarticletitle{Knowledge-Free Table Summarization}. In
  \bibinfo{booktitle}{\emph{Proc. of DWKD '13}}.
\newblock


\bibitem[\protect\citeauthoryear{Jain, Laha, Sankaranarayanan, Nema, Khapra,
  and Shetty}{Jain et~al\mbox{.}}{2018}]%
        {Jain:2018:AMH}
\bibfield{author}{\bibinfo{person}{Parag Jain}, \bibinfo{person}{Anirban Laha},
  \bibinfo{person}{Karthik Sankaranarayanan}, \bibinfo{person}{Preksha Nema},
  \bibinfo{person}{Mitesh~M. Khapra}, {and} \bibinfo{person}{Shreyas Shetty}.}
  \bibinfo{year}{2018}\natexlab{}.
\newblock \showarticletitle{A Mixed Hierarchical Attention Based
  Encoder-Decoder Approach for Standard Table Summarization}. In
  \bibinfo{booktitle}{\emph{Proc. of NCACL '18}}.
\newblock


\bibitem[\protect\citeauthoryear{Lo, Wu, and Yu}{Lo et~al\mbox{.}}{2000}]%
        {Ming:2000:TSA}
\bibfield{author}{\bibinfo{person}{Ming{-}Ling Lo}, \bibinfo{person}{Kun{-}Lung
  Wu}, {and} \bibinfo{person}{Philip~S. Yu}.} \bibinfo{year}{2000}\natexlab{}.
\newblock \showarticletitle{TabSum: {A} Flexible and Dynamic Table
  Summarization Approach}. In \bibinfo{booktitle}{\emph{Proc. ICDCS '00}}.
\newblock


\bibitem[\protect\citeauthoryear{Pasupat and Liang}{Pasupat and Liang}{2015}]%
        {pasupat2015compositional}
\bibfield{author}{\bibinfo{person}{Panupong Pasupat} {and}
  \bibinfo{person}{Percy Liang}.} \bibinfo{year}{2015}\natexlab{}.
\newblock \showarticletitle{Compositional Semantic Parsing on Semi-Structured
  Tables}. In \bibinfo{booktitle}{\emph{Proc. of ACL-IJCNLP '15}}.
\newblock


\bibitem[\protect\citeauthoryear{Radford, Wu, Child, Luan, Amodei, and
  Sutskever}{Radford et~al\mbox{.}}{2019}]%
        {radford2019language}
\bibfield{author}{\bibinfo{person}{Alec Radford}, \bibinfo{person}{Jeffrey Wu},
  \bibinfo{person}{Rewon Child}, \bibinfo{person}{David Luan},
  \bibinfo{person}{Dario Amodei}, {and} \bibinfo{person}{Ilya Sutskever}.}
  \bibinfo{year}{2019}\natexlab{}.
\newblock \bibinfo{title}{Language Models are Unsupervised Multitask Learners
  (Preprint)}.
\newblock
\newblock


\bibitem[\protect\citeauthoryear{Raffel, Shazeer, Roberts, Lee, Narang, Matena,
  Zhou, Li, and Liu}{Raffel et~al\mbox{.}}{2019}]%
        {raffel2019t5}
\bibfield{author}{\bibinfo{person}{Colin Raffel}, \bibinfo{person}{Noam
  Shazeer}, \bibinfo{person}{Adam Roberts}, \bibinfo{person}{Katherine Lee},
  \bibinfo{person}{Sharan Narang}, \bibinfo{person}{Michael Matena},
  \bibinfo{person}{Yanqi Zhou}, \bibinfo{person}{Wei Li}, {and}
  \bibinfo{person}{Peter~J. Liu}.} \bibinfo{year}{2019}\natexlab{}.
\newblock \bibinfo{title}{Exploring the Limits of Transfer Learning with a
  Unified Text-to-Text Transformer}.
\newblock
\newblock
\showeprint[arxiv]{1910.10683}


\bibitem[\protect\citeauthoryear{Sel\c{c}uk~Candan, Cao, Qi, and
  Sapino}{Sel\c{c}uk~Candan et~al\mbox{.}}{2009}]%
        {Selcuk:2009:ASS}
\bibfield{author}{\bibinfo{person}{K. Sel\c{c}uk~Candan},
  \bibinfo{person}{Huiping Cao}, \bibinfo{person}{Yan Qi}, {and}
  \bibinfo{person}{Maria~Luisa Sapino}.} \bibinfo{year}{2009}\natexlab{}.
\newblock \showarticletitle{AlphaSum: Size-Constrained Table Summarization
  Using Value Lattices}. In \bibinfo{booktitle}{\emph{Proc.~of EDBT'09}}.
\newblock


\bibitem[\protect\citeauthoryear{Vaswani, Shazeer, Parmar, Uszkoreit, Jones,
  Gomez, Kaiser, and Polosukhin}{Vaswani et~al\mbox{.}}{2017}]%
        {Vaswani:2017:AAY}
\bibfield{author}{\bibinfo{person}{Ashish Vaswani}, \bibinfo{person}{Noam
  Shazeer}, \bibinfo{person}{Niki Parmar}, \bibinfo{person}{Jakob Uszkoreit},
  \bibinfo{person}{Llion Jones}, \bibinfo{person}{Aidan~N. Gomez},
  \bibinfo{person}{Lukasz Kaiser}, {and} \bibinfo{person}{Illia Polosukhin}.}
  \bibinfo{year}{2017}\natexlab{}.
\newblock \showarticletitle{Attention is All You Need}. In
  \bibinfo{booktitle}{\emph{Proc. of NIPS '17}}.
\newblock


\bibitem[\protect\citeauthoryear{Wolf, Debut, Sanh, Chaumond, Delangue, Moi,
  Cistac, Rault, Louf, Funtowicz, and Brew}{Wolf et~al\mbox{.}}{2019}]%
        {Wolf2019HuggingFacesTS}
\bibfield{author}{\bibinfo{person}{Thomas Wolf}, \bibinfo{person}{Lysandre
  Debut}, \bibinfo{person}{Victor Sanh}, \bibinfo{person}{Julien Chaumond},
  \bibinfo{person}{Clement Delangue}, \bibinfo{person}{Anthony Moi},
  \bibinfo{person}{Pierric Cistac}, \bibinfo{person}{Tim Rault},
  \bibinfo{person}{Rémi Louf}, \bibinfo{person}{Morgan Funtowicz}, {and}
  \bibinfo{person}{Jamie Brew}.} \bibinfo{year}{2019}\natexlab{}.
\newblock \bibinfo{title}{HuggingFace's Transformers: State-of-the-art Natural
  Language Processing}.
\newblock
\newblock
\showeprint[arxiv]{1910.03771}


\bibitem[\protect\citeauthoryear{Zhang and Balog}{Zhang and Balog}{2020}]%
        {Zhang:2020:WTE}
\bibfield{author}{\bibinfo{person}{Shuo Zhang} {and} \bibinfo{person}{Krisztian
  Balog}.} \bibinfo{year}{2020}\natexlab{}.
\newblock \showarticletitle{Web Table Extraction, Retrieval, and Augmentation:
  A Survey}.
\newblock \bibinfo{journal}{\emph{ACM Trans. Intell. Syst. Technol.}}
  (\bibinfo{year}{2020}).
\newblock


\end{thebibliography}
